\documentclass[epj]{webofc}
\usepackage[varg]{txfonts}   
\woctitle{MENU 2013}
\begin{document}
\title{Hadron form factors and large-$N_c$ phenomenology}
%
%

\author{Pere Masjuan\inst{1}\fnsep\thanks{Speaker. \email{masjuan@kph.uni-mainz.de.} Supported by MICINN of Spain (FPA2010-16696, FIS2011-24249), Junta de Andaluc\'ia (FQM 225), by the Deutsche Forschungsgemeinschaft DFG through the Collaborative Research Center ``The Low-Energy Frontier of the Standard Model" (SFB 1044), by Polish NCN grant DEC-2011/01/B/ST2/03915. P.M. is partially supported by the Jefferson Science Associates, LLC. } \and
        Enrique Ruiz Arriola\inst{2} \and
        Wojciech Broniowski\inst{3}        
}

\institute{PRISMA Cluster of Excellence, Institut f\"ur Kernphysik, Johannes Gutenberg-Universit\"at, D-55099 Mainz, Germany
\and
         Departamento de F\'isica At\'omica, Molecular y Nuclear and Instituto Carlos I de F\'isica Te\'orica y Computacional, Universidad de Granada, E-18071 Granada, Spain
\and
           The H. Niewodnicza\'nski Institute of Nuclear Physics PAN, PL-31342 Krak\'ow and Institute of Physics, Jan Kochanowski University, PL-25406 Kielce, Poland
          }

\abstract{%
The \textit{half width rule} provides a way to consider $1/N_c$ corrections to hadronic models containing resonances. Consequences of such ideas for hadron form factors and Regge trajectories are explored, with special emphasis on the possibility to describe the spectrum of light and heavy unflavored vector mesons in a universal way.
}
\maketitle
\vspace{-0.4cm}
\section{Introduction}
\label{intro}

The rigorous quantum-mechanical definition of a resonance with given quantum numbers corresponds to a pole in the second Riemann sheet in the (analytically continued) partial-wave amplitude of the considered scattering channel. This definition becomes independent of the background, whereas the corresponding residue provides the coupling to produce that resonance in the given process. Particle Data Tables~\cite{PDG} rarely quote such pole positions for practical reasons and rely on definitions such as a pole in the K-matrix, the Breit-Wigner resonance, the location of a maximum in the speed plot, the time delay, etc. Complex energies cannot be measured and an analytic continuation to the complex plane is required. If the amplitude on the real axis is just approximated or modeled, the analytic continuation might increasingly amplifie the uncertainty with the resonance width. Different definitions would yield different mass parameters, although they should converge in the narrow-resonance limit. 

Our observation is that since in the large-$N_c$ limit $\Gamma/M = {\cal O}(N_c^{-1} )$~\cite{largeNc}, the maximum level of discrepancy in quoting resonance mass parameters $M$ should be compatible with its own width $\Gamma$, i.e., in the interval $ M \pm \Gamma/2$, the \textit{half-width rule}. In this presentation, we discus the implications of such an idea for hadron form factors and linear Regge trajectories~\cite{Masjuan:2012sk,Masjuan:2012gc}. Applications to the hadronic density of states, the anomalous magnetic moment of the muon, and the analysis of MiniBooNE data are discussed in Ref.~\cite{Arriola:2012vk}.

\section{Hadron form factors}
\label{sec1}

\begin{figure}
\centering
\includegraphics[width=7cm]{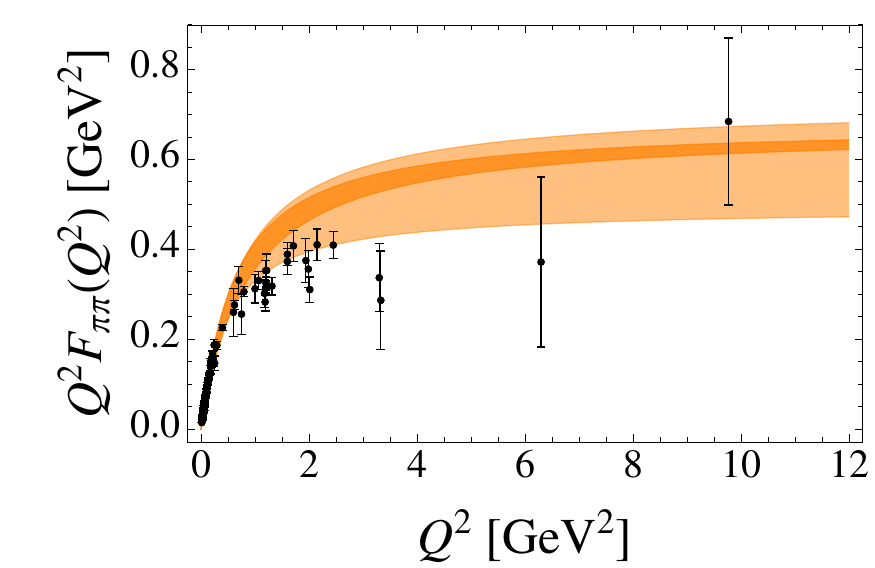}
\includegraphics[width=7cm]{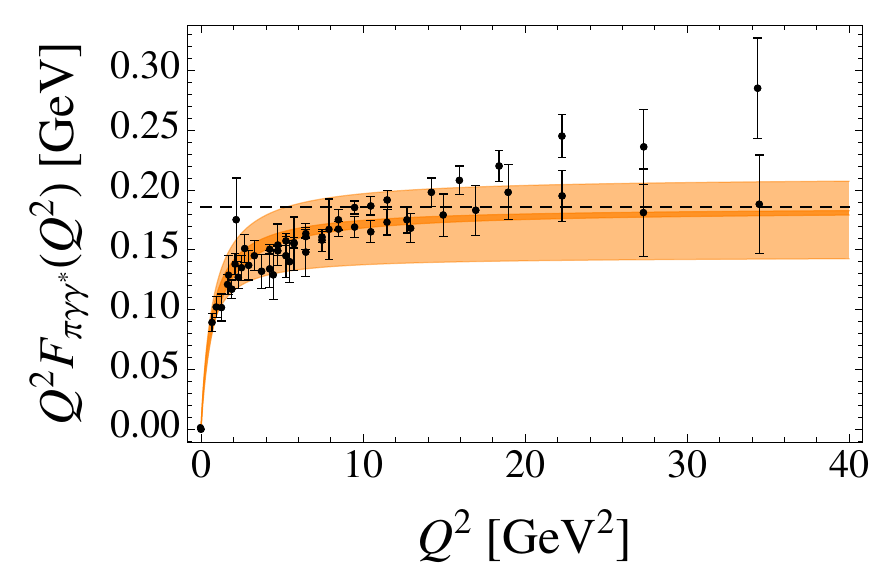}
\caption{Electromagnetic form factor $F(Q^2)$ (left panel) and transition form factor $F_{\pi^0 \gamma}(Q^2)=F(Q^2)/(4\pi^2 f_{\pi})$ (right panel) of the pion parameterized with the meson dominance ansatz~(\ref{vmd}) with error bands given by the HWR. Light orange bands correspond to the monopole ansatz and the darker inner bands the two resonance ansatz.}
\label{fig1}       
\end{figure}

The mass shift and the width of a resonance are parametrically of the same order in the large $N_c$. As an example, let us take the two-point Green Functions propagator $D(s)=(s-m_0^2- \Sigma(s))^{-1}$, with $m_0$ the bare mass and $\Sigma(s)$ the self-energy. The resonance pole $s_R=m_R^2- i m_R \Gamma_R$ is defined as $s_R - m_0^2-\Sigma(s_R)=0$. In the large-$N_c$ limit, one can take a perturbative expansion of the pole position:
$s_R = m_0^2 + {\cal O}(N_c^{-1})$ since the mesonic loops (i.e, the pion loop giving width to the $\rho$-meson, for example) are suppressed in large-$N_c$~\cite{largeNc}. The next order in $1/N_c$ would correspond to including such loops $s_R = m_0^2 + 2m_0 \Delta m_R - i m_0\Gamma_R + {\cal O}(N_c^{-2})$. The imaginary part of the pole position, related to the width of the particle, is parametrically identical to the mass shift of the real part of the pole position. Calculations made with resonance models based in the large-$N_c$ limit that use the real part of the PDG value for calculations are making an error given parametrically by the width of the particle. The half-width rule method (HWR) represents an estimation of such an error.

When considering hadron form factors within the HWR such as the ones in Fig.~\ref{fig1}
\begin{equation}\label{vmd}
F(Q^2)=\frac{M_V^2}{M_V^2+Q^2}\, \textrm{  (one pole ansatz) } , \quad 	F(Q^2)=\frac{M_V^2 M_V^{'2}+ A Q^2}{(M_V^2+Q^2)(M_V^{'2}+Q^2)}\, \textrm{  (two poles ansatz)} ,
\end{equation}
\noindent
the following is assumed: hadronic form factors in the spacelike region are dominated by mesonic states with the relevant quantum numbers; the high-energy behavior is given by perturbative QCD [which determines the parameter $A$ in~(\ref{vmd})], and the number of mesons is taken to be minimal to satisfy these conditions; errors in the meson-dominated form factors are estimated by the HWR by treating masses as random variables distributed with the dispersion given by the width. $1/N_c$ corrections are linked with phenomenological predictions at $N_c=3$ through the errors provided by the HWR.

We find that within the HWR nucleon form factors usually parameterized as dipoles can be described as a product  of two monopoles with overlapping resonance masses.

\section{Linear Regge trajectories}

\textit{Radial} linear Regge trajectories ($M_n^2 = M_0^2 + \mu^2 n$)~\cite{Kang} are a consequence of confinement in quark models as a generalization of the \textit{angular-momentum} linear Regge trajectories ($M_J^2 = M_0^2 + \beta^2 J$)  by Chew and Frautschi~\cite{Chew}.  In Ref.~\cite{Anisovich:2000kxa} it was suggested that the light-quark meson states grouped into radial Regge trajectories would have the slope $\mu^2=1.25(15)$~GeV$^2$, where the error was estimated as the spread of the values for each meson-family considered ($\rho,\pi,\eta,a,f$).
In Ref.~\cite{Masjuan:2012gc} we reanalyzed the radial and angular-momentum Regge trajectories with the updated list of the light unflavored mesons from the Particle Data Tables~\cite{PDG}, using the HWR~\cite{Masjuan:2012gc,RuizArriola:2010fj} for the fits and error estimates, i.e, the half-width squared as a weight for each resonance, through a $\chi^2$ minimization fit with $\chi^2=\sum_n \left( \frac{M_n^2-M_{n,exp}^2}{\Gamma_n M_n}\right)^2$. 

The fit to all the light unflavored meson families with linear trajectories using the HWR yields $\mu^2=1.35(4)$~GeV$^2$ as the weighted averaged result for the slope of radial trajectories, and $\beta^2=1.16(4)$~GeV$^2$ as the weighted average for the slope of the angular-momentum trajectories~\cite{Masjuan:2012gc}. We also considered a joint fit with the formula $M^2_X(n,J)=M^2_X(0,0)+ n \mu^2 + J \beta^2$, with the result $M^2_X(n,J)=(-1.25(4)+ 1.38(4) n  + 1.12(4)J)$~GeV$^2$, which means a difference between the radial and the angular-momentum slopes at a statistically significant level of 4.5 standard deviations. Our analysis, in agreement with relativistic quark models~\cite{RelModel}, are in clear contradiction to the assumption of universality of radial and angular-momentum slopes $M^2 \sim n+J$ ~\cite{Afonin:2006vi}.

As an extension of Ref.~\cite{Masjuan:2012gc}, we presented in Ref.~\cite{Masjuan:2012yw} a study of both radial and angular-momentum trajectories for the kaon sector. The radial fit yields $\mu^2_K=1.22(21)$~GeV$^2$ and $1.12(21)$~GeV$^2$ for $K$ and $K^*$, respectively, while the angular-momentum fit returns $\beta^2_K=1.36(6)$~GeV$^2$ and $1.19(7)$~GeV$^2$ for $K$ and $K^*$, respectively. Only trajectories containing three or more states are considered. The Kaon, as a Goldstone Boson, should not be included in the trajectories.

\subsection{Universal radial Regge trajectory for unflavored vector mesons}

\begin{figure}
\centering
\includegraphics[width=6cm]{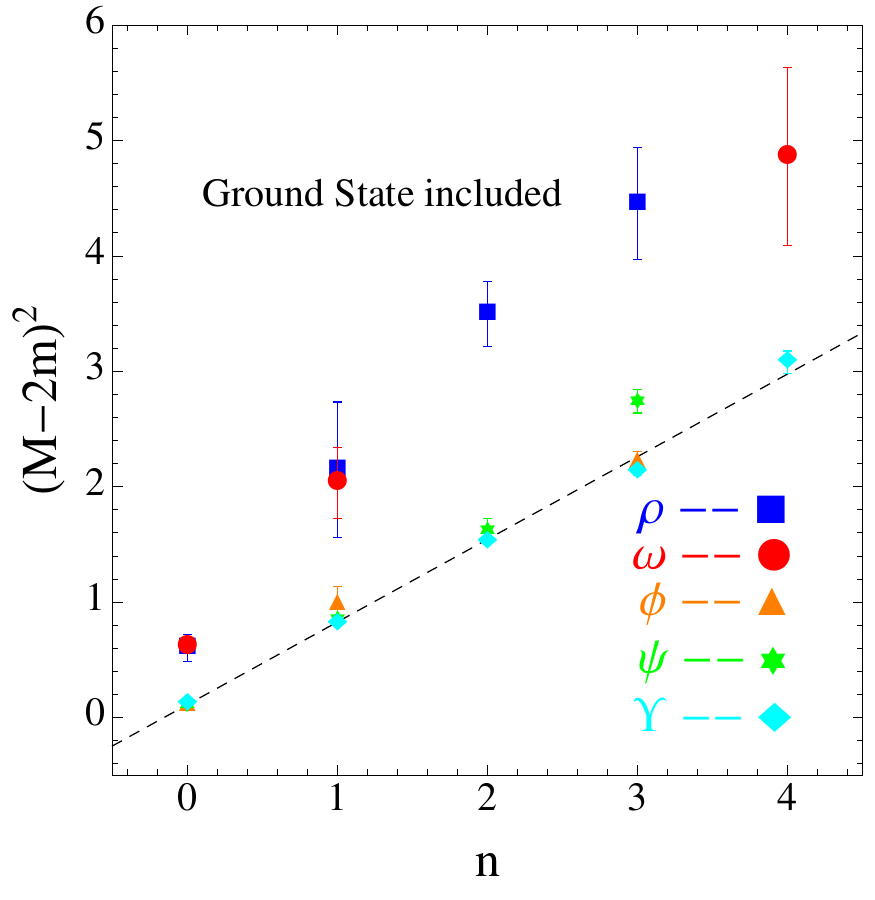}
\includegraphics[width=6cm]{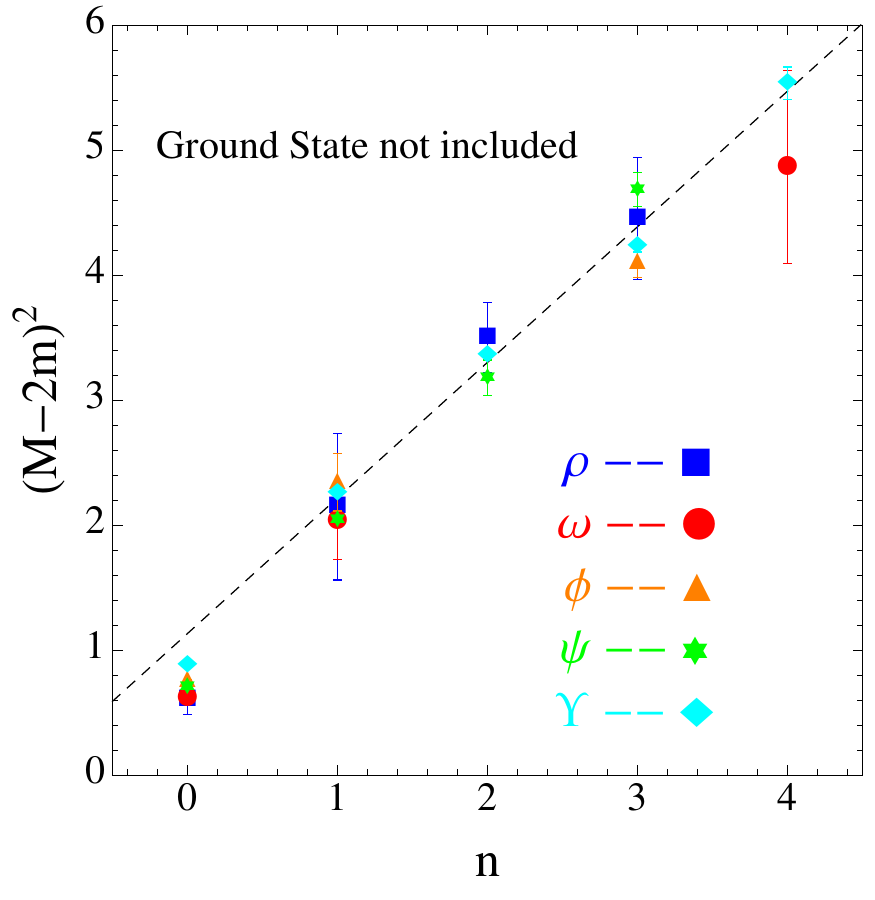}
\caption{Radial Regge trajectory for the unflavored light and heavy vector mesons. Left panel includes the ground state while right panel does not include the ground state.}
\label{fig2}       
\end{figure}

Afonin and Pusenkov proposed in Ref.~\cite{Afonin:2013hla} that the spectrum of light and heavy unflavored vector mesons can be parameterized in  a universal way by a modification of the usual radial Regge trajectory

\begin{equation}\label{equni}
(M_n-2m)^2=a(n+b)
\end{equation}
\noindent
where $m$ is related to the quark mass flavor, $a$ would be the universal slope related with the QCD hadronic scale, and $b$ the universal offset parameter, related to the mass gap in QCD. They remarkably concluded that, indeed, such universal trajectory exists after fitting the $\omega,\phi,\Psi$, and $\Upsilon$ mesons and obtained $a=1.10, b=0.57$ (considering $m_u=m_d=0$). In their analysis, however,  two shortcomings are found: the lack of error analysis and the inclusion of the ground state on the linear trajectory.

This scenario is ideal for an application of our HWR method. Using Eq.~(\ref{equni}) together with the HWR and including the $\rho$-meson states  on top of the $\omega,\phi,\Psi$, and $\Upsilon$ states we provide a fit to the universal radial trajectory.  Figure~\ref{fig2} is the outcome of our method  and the results of the fits are collected in Table~\ref{tab1}. Including the ground state yields a very bad fit (left panel in Fig.~\ref{fig2}) since the trajectory is dominated by the $\Upsilon$ family and $\rho$ and $\omega$ mesons lie far above the line. Including the ground state does not result in the desired universality.  However, keeping apart the ground state (right panel in Fig.~\ref{fig2}), results in a reasonably good fit ($\chi^2 \sim 0.1$) and yields $a=1.03(19)$~GeV$^2$ and $b=0.9(5)$~GeV$^2$ in agreement with the results of the global analysis of the radial Regge trajectories. The results for the constituent quark masses are impressively similar than those of the Particle Data Tables, i.e, $m_c=1.275(25)$~GeV and $m_b=4.18(3)$~GeV~\cite{PDG}.

\begin{table}
\centering
\caption{Fitted parameters for the universal Regge trajectory Eq.~(\ref{equni}) including (second column) or not (third column) the ground state in the fit. The mass of the light quarks are $m_u=m_d=0$ MeV.}
\label{tab1}       
\begin{tabular}{ccc}
& Ground state included & Ground state {\bf not} included\\
\hline
a & 0.716(2) & 1.03(19)\\
b & 0.153(1) & 0.93(50)\\
\hline
$m_s$ & $0.344(1)$ GeV & $0.12(67) $ GeV \\
$m_c$ & $1.383(1)$ GeV & $1.17(12) $ GeV \\
$m_b$ & $4.561(1)$ GeV & $4.30(10) $ GeV \\
\hline
\end{tabular}
\end{table}

\end{document}